\documentclass[12pt,preprint]{aastex}

\newcommand\xte{{\it RXTE\/}}

\newcommand\chandra{{\it Chandra}}

\def\snr{\hbox{G54.1$+$0.3}}
\def\psr{\hbox{PSR~J1930$+$1852}}

\def\simlt{\mathrel{\hbox{\rlap{\hbox{\lower4pt\hbox{$\sim$}}}\hbox{$<$}}}}
\def\simgt{\mathrel{\hbox{\rlap{\hbox{\lower4pt\hbox{$\sim$}}}\hbox{$>$}}}}

\slugcomment{draft version of Feb. 19 2007}

\shorttitle{X-ray Observations of \psr}
\shortauthors{Lu et al.}

\begin{document}


\title{X-ray Timing Observations of \psr\ in the Crab-like 
SNR \snr}


\author{Fangjun Lu\altaffilmark{1,2}, Q.Daniel Wang\altaffilmark{2}, E. V. Gotthelf\altaffilmark{3}, and Jinlu Qu\altaffilmark{1}}


\altaffiltext{1}{Laboratory of Particle Astrophysics, 
Institute of High Energy Physics, CAS, Beijing 100039, P.R. China; 
lufj@mail.ihep.ac.cn; qujl@mail.ihep.ac.cn}

\altaffiltext{2}{Astronomy Department, University of Massachusetts,
    Amherst, MA 01003; wqd@astro.umass.edu}

\altaffiltext{3}{Columbia Astrophysics Laboratory, Columbia University, 550 West 120th Street, New York, NY 10027; eric@astro.columbia.edu}

\begin{abstract}
We present new X-ray timing and spectral observations of \psr, the
young energetic pulsar at the center of the non-thermal supernova
remnant \snr. Using data obtained with the {\it Rossi} X-ray Timing
Explorer (\xte) and \chandra\ X-ray observatories we have derived an
updated timing ephemeris of the 136~ms pulsar spanning 6 years. During
this interval, however, the period evolution shows significant
variability from the best fit constant spin-down rate of $\dot P =
7.5112(6) \times 10^{-13}$~s~s$^{-1}$, suggesting strong timing noise and/or
glitch activity.  The X-ray emission is highly pulsed ($71\pm5\%$
modulation) and is characterized by an asymmetric, broad profile
($\sim 70\%$ duty cycle) which is nearly twice the radio width. The
spectrum of the pulsed emission is well fitted with an absorbed power
law of photon index $\Gamma = 1.2\pm0.2$; this is marginally harder
than that of the unpulsed component.  The total $2-10$~keV flux of the
pulsar is $1.7 \times 10^{-12}$~erg~cm$^{-2}$~s$^{-1}$. These results
confirm \psr\ as a typical Crab-like pulsar.
\end{abstract}


\section{Introduction}

Young rotation-powered pulsars typically radiate a
large fraction of their spin-down energy at X-ray energies.
Observations in this band are thus important to the study of 
the spin-down evolution of such pulsars and their emission mechanism(s).
The study also helps to understand the mechanical energy output of the
pulsars into their surroundings, manifested as pulsar wind
nebulae (PWNe). To this end, one needs to monitor the spin-down at
various evolutionary stages of young pulsars and to measure their
energy spectra, both pulsed and unpulsed, with various viewing
angles.  However, only a dozen or so of young pulsars with PWNe have been
identified and studied in detailed so far.

The recently discovered 136~ms pulsar \psr\ at the center of the
supernova remnant (SNR) \snr\ is the latest example of a Crab-like
pulsar (Camilo et al. 2002).  Known as the ``Bulls-Eye'' pulsar, \psr\
is surrounded by a bright symmetric ring of emission (Lu et al. 2002)
similar to the toroidal and jet-like structure associated with the
Crab pulsar, but viewed nearly face-on. Based on the initial timing
parameters, \psr\ is the eighth most energetic pulsar known, with a
rotational energy loss rate of $\dot E = 1.2 \times
10^{37}$~erg~s$^{-1}$, well above the empirical threshold for
generating a bright pulsar wind nebula ($\dot E \simgt 4 \times
10^{36}$~erg~s$^{-1}$, Gotthelf 2004).
Such young pulsars are often embedded in observable shell-type
remnant which have yet to dissipate. However, like the Crab, G54.1+0.3
lacks evidence for a thermal remnant in any waveband (Lu et
al. 2002). Most likely, the SN ejecta in these two remnants are still
expanding into a very low density medium.

In this paper we present the first dedicated X-ray timing and
spectral follow-up observations of \psr\ since discovery. Previous
X-ray results were based on archival data of limited quality. We use
the new data to characterize the pulse shape and energy spectrum and
provide a long term ephemeris.  Throughout the paper, the
uncertainties (statistical fluctuation only) are quoted at the 68\%
confidence level.
 
\section{Observations and Data Analysis}

The pulsar \psr\ was observed twice with \xte\ on 2002 September 12 --
14 and on 2002 December 23 -- 25 using a combination of event and
instrument modes. For consistency, we analyze the data taken with the
proportional counter array (PCA) in the Good~Xenon mode. PCA has a 
field of view of 1$\degr$ (FWHM), total collecting area 
of about 6500 cm$^2$, time resolution of 1 $\mu$s, and spectral 
resolution of $\leq18\%$ at 6 keV. The data are
reduced and analyzed using the {\sl ftools} software package version
v5.2. We filter the data using the standard \xte\ criteria,
selecting time intervals for which parameters ${\tt Elevation\_Angles}
<10\degr$, ${\tt Time\_Since\_SAA} \geq 30$ min, ${\tt
Pointing\_Offsets} < 0.02 \degr$, and the background 
electron rate ${\tt Electron2} < 0.1$.  The effective exposure 
time after this filtering is 31.7~ks and
41.7~ks for the September and December observations.  Since the background
of \xte\ is high and the spectral resolution is relatively low,
the \xte\ data
is used herein exclusively for timing analysis, selecting photons
detected from PCA PHA channels $0-35$ ($\sim 2-15$~keV). This results
in a total of $\sim 1$ and $\sim 1.6$ million counts in the
two observations for the subsequent analysis. The photon arrival 
times are corrected
to the Solar system barycenter, based on the DE200 Solar ephemeris time
system and the \chandra\ J2000 coordinates of J193030.13+185214.1 
(Lu et al. 2002).

SNR \snr\ was also observed with \chandra\ on 2003 June 30 for a total
of 58.4 ks. The pulsar was placed at the aim-point of the
front-illuminated ACIS-I detector. The CCD chip I3 was operated in
continuous-clocking mode (CC-mode), providing a time resolution of
2.85~ms and an one-dimensional imaging, in which the \hbox{2-D} CCD image is
integrated along the column direction in each CCD readout cycle. The
photon arrival times are post-processed to account for the spacecraft 
dithering and SIM motion prior to the barycenter correction. 
The spectral data are corrected for the effects of CTI (Charge
Transfer Inefficiency). However, the spectral gain is not well
calibrated in the CC-mode, requiring adjustment in the fitting
process (details are given in \S 3). Spectral response matrices are
generated for the ACIS-I aimpoint, the location of the pulsar in this
observation.  After filtering the data using the standard criteria,
the remaining effective exposure is 57.2~ks.  Reduction and analysis
of the \chandra\ data are all based on the standard software package
{\sl CIAO} (v3.2) and {\sl CALDB} (v3.0.0).

Figure~1 presents the geometry of the CC-mode observation overlaid on
an archival \chandra\ X-ray image of SNR~G54.1+0.3.  The CCD image is
summed along the dimension perpendicular to the marked line which is
orientated with a position angle $P.A. = 19\degr$ East of North. The
count distribution along this dimension is shown in Figure~2. The
central peak corresponds to the presence of the pulsar, which
significantly contributes to the six adjacent pixels, as denoted by
the upper horizontal bar. The neighboring four pixels (two on each
side of the pulsar region), marked by the two lower horizontal bars,
show the nearly same intensity level in the ACIS-I3 image-mode data with the
pulsar excised. We therefore select counts
falling in the inner six pixels for both our pulsar timing and
spectral analysis of the pulsar, while those counts in the outer four
pixels are used to estimate the background from the surrounding
nebula.

\begin{figure}
\plotone{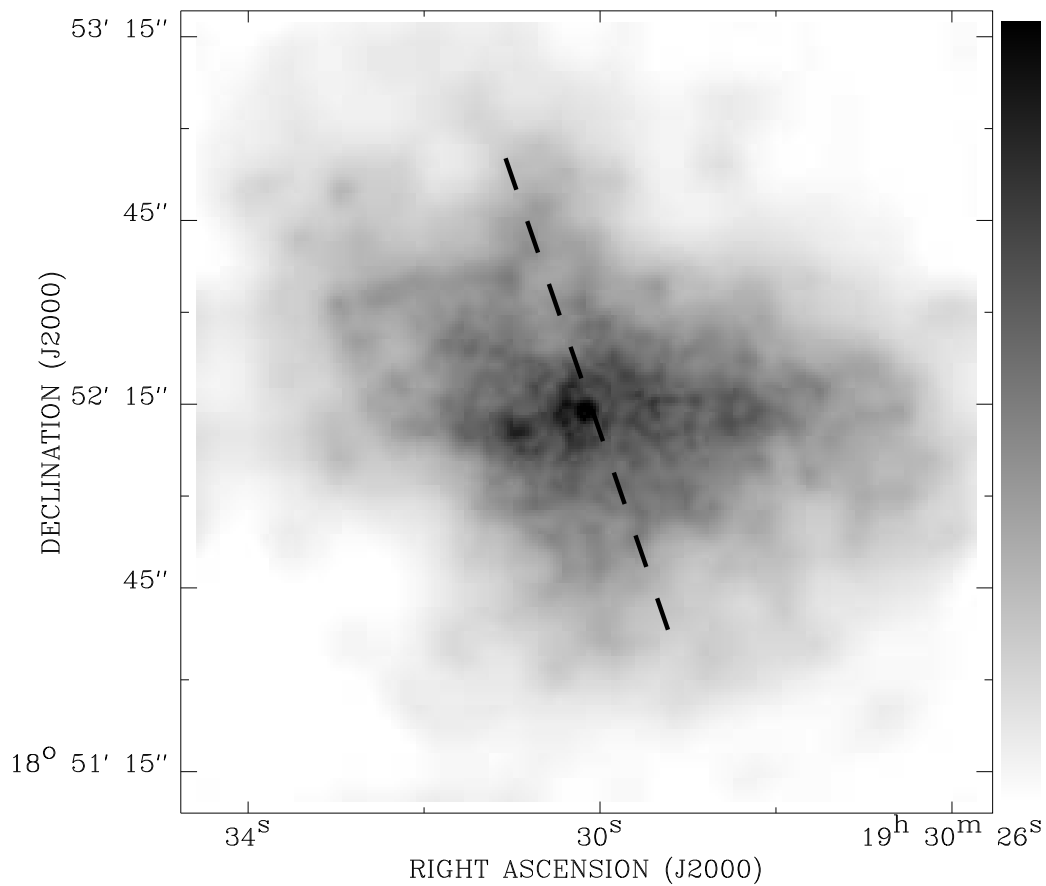}
\caption{Geometry of the \chandra\ ACIS-I3 CCD continuous-clocking
(CC-mode) observation of \psr\ presented herein. The dashed line gives
the orientation of the CC-mode observation shown overlaid on an
archival \chandra\ broadband ($0.3-10$~keV) X-ray image of SNR G54.1+0.3.
\label{fig1}}
\end{figure}  

\begin{figure}
\plotone{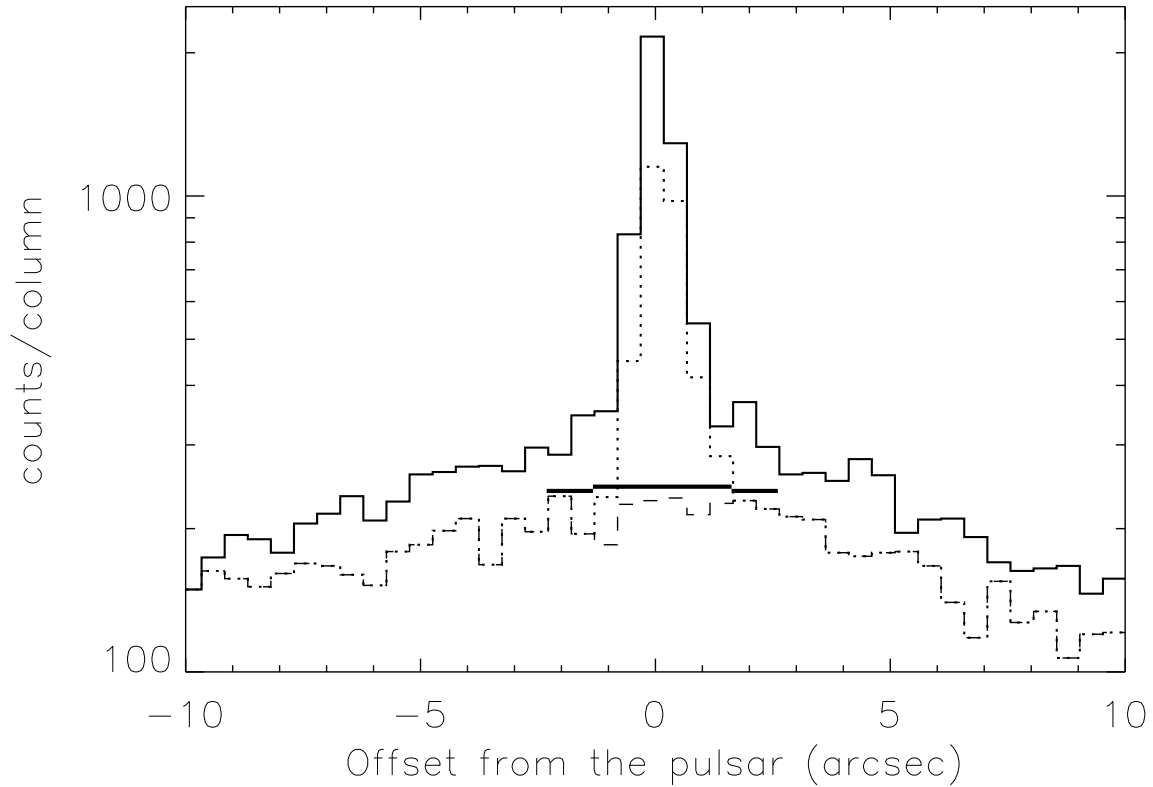}
\caption{Source and background region determination for the \chandra\
CC-mode observation of \psr. The \hbox{1-D} count distribution of
SNR~G54.1+0.3 as observed in CC-mode using ACIS-S3 ({\it solid line}),
compared with the distributions constructed from the collapsed
\chandra\ ACIS-I3 image-mode data, with ({\it dashed line}) and
without ({\it dotted}) the pulsar excised. All data is restricted to
the $0.3-10$~keV energy band. The on-pulsar (central) thick horizontal bar
denotes the 6 pixels that contains significant pulsar emission, while
the two adjacent off-pulsar (outer) thick horizontal bars mark the pixels
that are used to estimate the local nebula background (see \S2 for
details).
\label{fig2}}
\end{figure}

\section{Results}

\subsection{Pulsar Timing}

For each observation, we search for the periodic signal of \psr\ by
folding events around the period extrapolated from the early radio
ephemeris of Camilo et al (2002). For each period folding with a
period $P$, a $\chi^2$ is calculated from the fit to the pulse profile
with a constant count rate. The null hypothesis of no periodic signal can be
ruled out when a significant peak is seen in the resultant
``periodogram'' ($\chi^2$ vs. $P$), which is the case for each of the
X-ray observations at a high confidence ($\chi^2 > 300$ for 10 phase
bins). We further fit the peak shape with a Gaussian profile to
maximize the accuracy of our pulsar period determination (Figure~3).
The centroid of this Gaussian is then taken as the best estimate of
the pulsar period. The light curves derived of the \xte\ and \chandra\
observations folded at the measured periods are shown in Figures.~4--5.

To estimate the uncertainties in the period measurements, we use the
bootstrap technique of Diaconis \& Efron (1983). This is done as the
following: (1) constructing a new data set of the same total 
number of counts by re-sampling with replacement from the observed events;
(2) determining the period with this re-sampled data set in the exactly 
same way as with the original data; (3) repeating the above two
steps for 500 times to produce a period distribution; (4) Using the
dispersion of this distribution as an estimate of the 1$\sigma$ period
uncertainty.  The distributions produced for the three observations
are shown in the right column of Figure~3, while the estimated
uncertainties are included in Table~1.

\begin{figure}[b]
\plotone{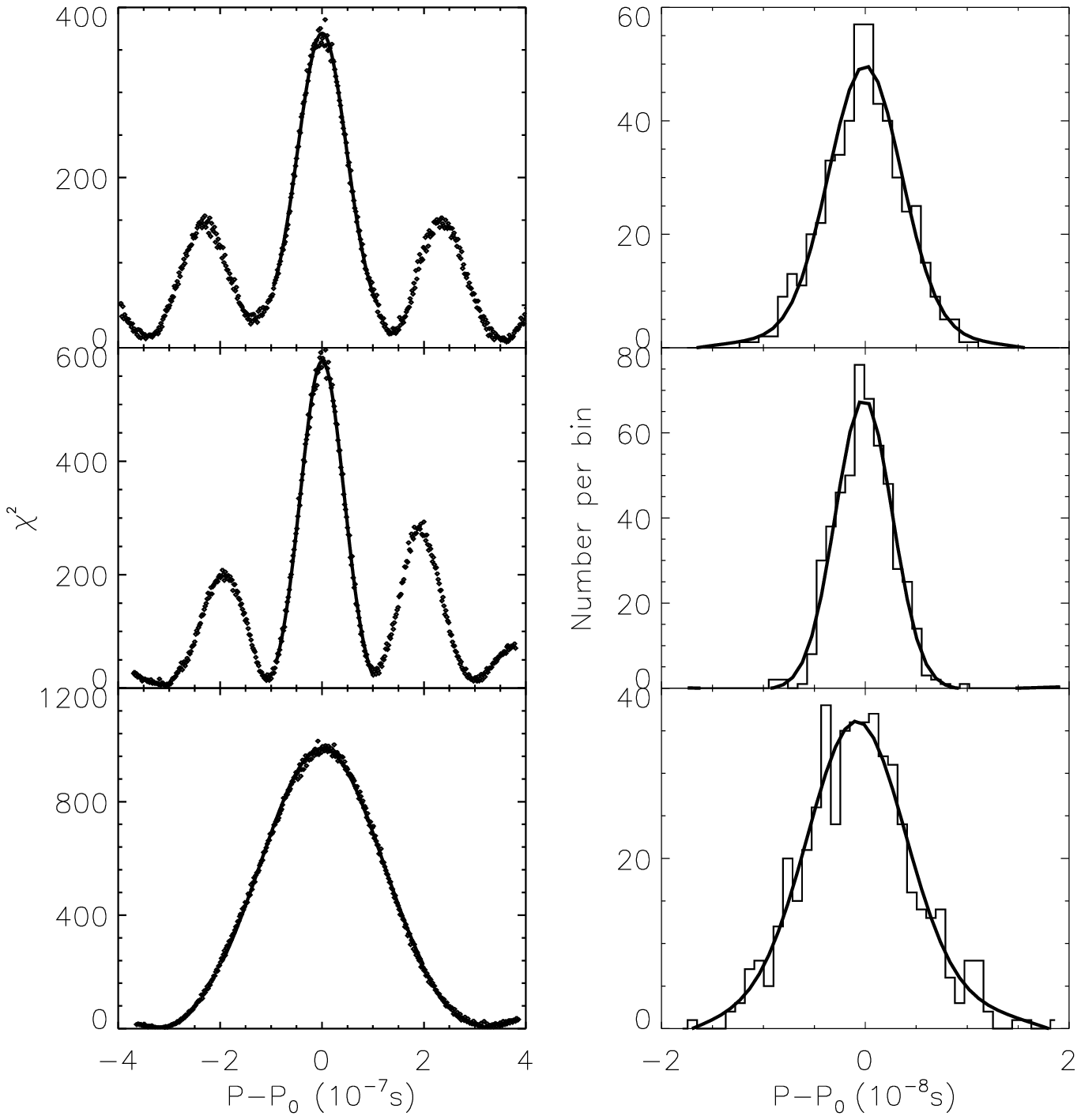}
\caption{Period and period uncertainty of \psr\ at three epochs. {\it
Left} -- The periodograms of PSR J1930+1852 constructed from the
September 2002, December 2002, and June 2003 observations and together
with the respective best-fit Gaussian profiles for the central
peaks. {\it Right} -- The
distribution of the 500 periods from the bootstrapped data for each
observations. The Gaussian $1\sigma$ width gives an estimate of the period
uncertainty. The P$_0$ values are given in Table 1.
\label{fig3}} 
\end{figure}

\begin{figure}[b]
\plotone{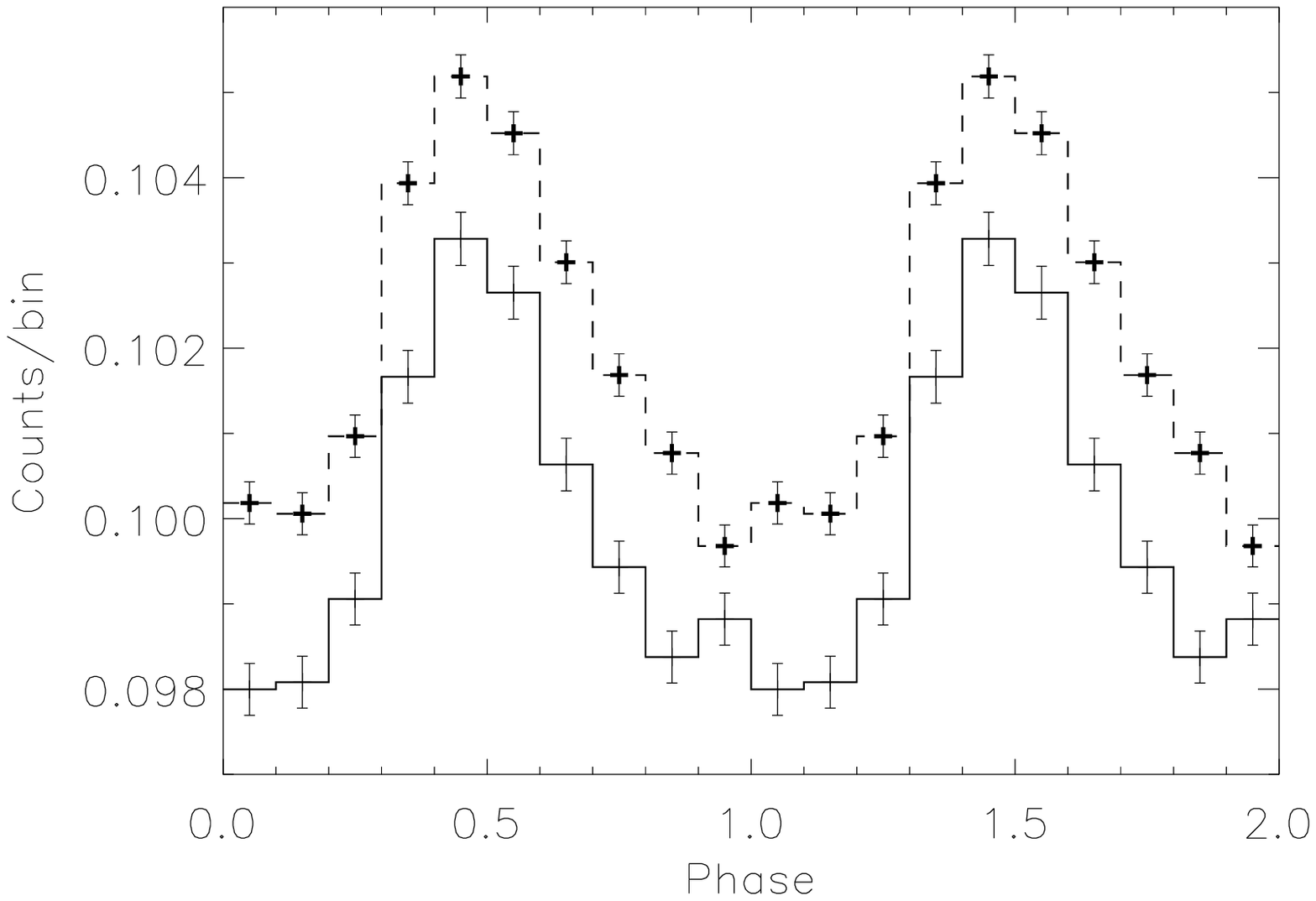}
\caption{The pulse shape of \psr\ in the $2 - 15$ keV band as
obtained with \xte\ on 2002 September 12 (solid) and December
23 (dashed). Phase zero is arbitrary; two cycles are shown for clarity.
The December light curve is shifted upward by 0.002.
\label{fig4}}
\end{figure}

To compute the pulsed fraction of the X-ray emission from 
\psr, we used the \chandra\
observation.  We extracted a total of 5506 counts
in the $0.3-10$ keV band from the on-pulsar pixels of the \hbox{1-D} count 
distribution (the solid curve in Figure~2).
After subtracting the local nebular contribution estimated
from the neighboring off-pulsar pixels, the remaining 
3560$\pm$92 counts are considered as the net total emission from the pulsar. 
This emission can be further divided into
the pulsed  and persistent components. To determine the  persistent component,
we construct a \hbox{1-D} distribution of the persistent emission from 
the off-pulse counts, defined to be in the phase interval $0.1 - 0.3$ (Figure~5).
The same on-pulsar pixels as shown in Figure~2 now contain a total of 598 counts,
Corrected for the off-pulse phase fraction (1/5), the total 
number of  persistent counts
over the entire phase is then 598$\times$5.  
Therefore, the net number of the pulsed counts is 
(5506-598$\times$5)=2516$\pm$143. This results in a pulsed fraction 
of $f_p \equiv ( {\rm pulsed/total\ counts}) = 71\pm5\%$.

\begin{figure}
\plotone{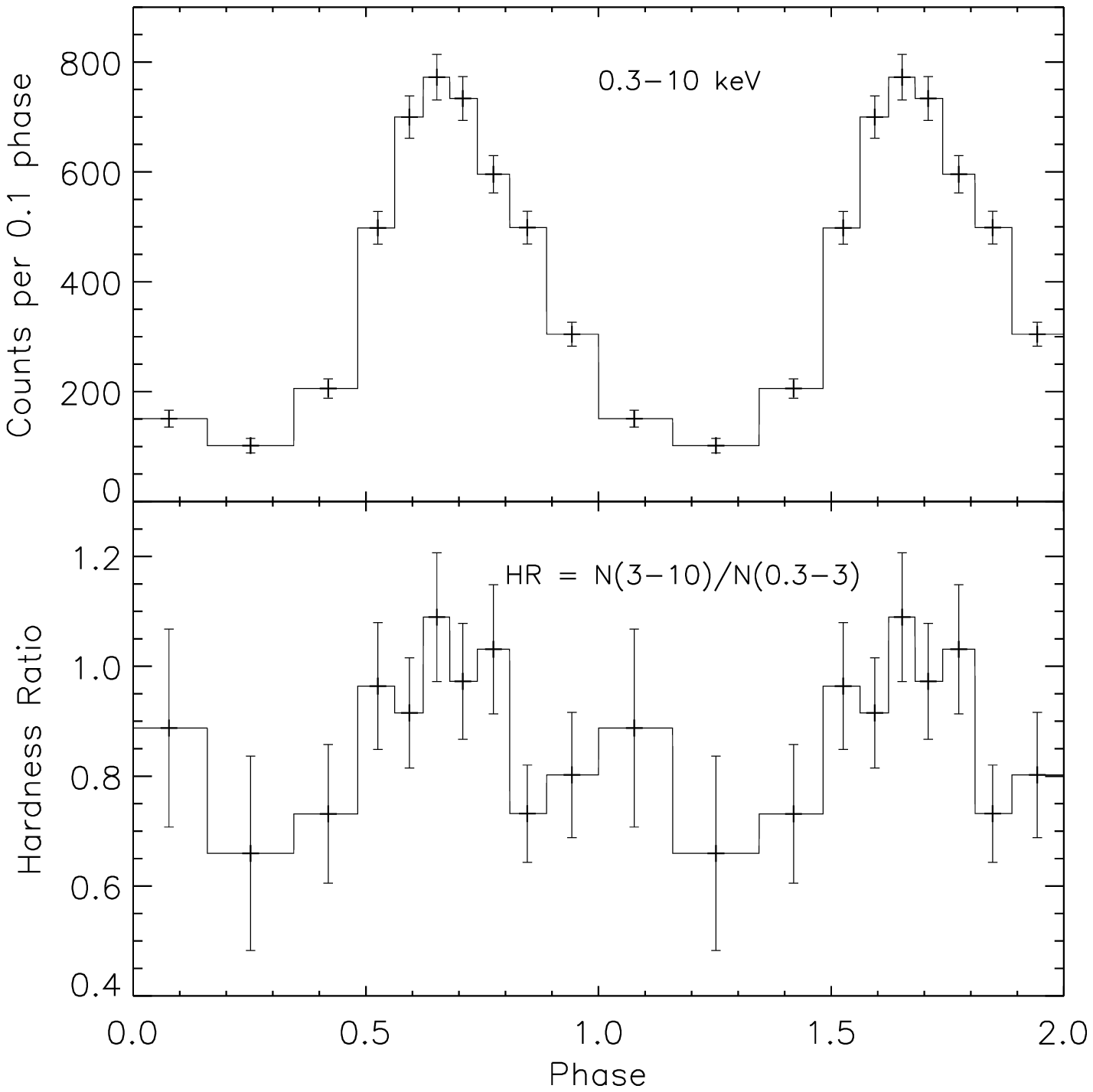}
\caption{The pulse shape and its hardness ratio of \psr\  in the 
$0.3-10$~keV band as measured with \chandra\ on 2002 June 30.  The
pulse shape ({\it Top Panel}) is folded at the period given in Table~1
and the phase bin size is chosen so that each bin contains almost the same
counts. The hardness ratio {(\it Bottom Panel)} is as defined in the
text (\S3.2); the background, as defined in \S2, has been subtracted.
\label{fig5}}
\end{figure}

\begin{deluxetable}{lccc}
\tablewidth{0pt}
\tablecaption{Timing Results for PSR J1930+1852}
\tablehead{
\colhead{Date} & \colhead{Obs. Type} & \colhead{Epoch} &  \colhead{Period} \\
\colhead{(UT)} & &\colhead{(MJD[TDB])} &  \colhead{(s)}}
\startdata
1997 Apr 27 &ASCA & 50566  & 0.13674374(5){\tablenotemark{a}} \\
2002 Jan 17 &Radio & 52280  & 0.136855046957(9){\tablenotemark{a}} \\

2002 Sep 12 &RXTE  & 52530  & 0.136871312(4) \\
2002 Dec 23 &RXTE  &  52632  & 0.136877919(3) \\
2003 Jun 30 &Chandra &52820  &  0.136890130(5) \\

\enddata
\tablenotetext{a}{\footnotesize Taken from Camilo et al. (2002)}
\label{timetable}
\end{deluxetable}

\subsection{Pulsed Emission Spectral Characteristics}

To check for phase-dependent spectral variations across the pulse
profile we compute the hardness ratio in each phase bin, 
defined as
${\rm HR} = N_h/N_s$, where $N_s$ and $N_h$ are the counts selected
from the $0.3-3$~keV and $3-10$~keV energy bands, respectively.  The
pulsar counts (pulsed and unpulsed) are extracted from the 6 pixel
source region as discussed in \S2 and the background from the
neighboring 4 pixels.  The calculated  ${\rm HR}$ is 
shown in the lower panel of Figure~5. Fitting these ${\rm HR}$ 
data points  assuming a constant  ${\rm HR}$ value resulted 
in a $\chi^2$ of 17.94 for 9 degrees of freedom, which means 
that the hardness ratio changes with phase at a confidence level 
of 96.4$\%$. Further more, it appears that the ${\rm HR}$ values of the 
on pulse emission are higher than those of the off-pulse 
emission. In order to quantify this, we computed the mean ${\rm HR}$
for the off-pulse emission (bins 1, 2 3 and 10 in the panel) 
as ${\rm HR} = 0.77\pm0.08$ and the on-pulse bins (4 to 9) as
${\rm HR} = 0.95\pm0.04$.  Therefore, the on-pulse emission is harder than
the off-pulse emission at a confidence level of $\sim 2 \sigma$, or 98$\%$.

Next, we study the \chandra\ spectrum of PSR J1930+1852 using the same
sources and background counts as extracted above. For the pulsed
spectrum, the phase width corrected off-pulse counts are subtracted
from the on-pulse counts in each spectral bin. Figure~6a presents the
best fit absorbed power-law model using the standard response
matrix. Although the overall $\chi^2$ is acceptable (34.4 for 35
degree-of-freedom), the residuals to this fit display characteristic
feature, indicating that the gain of the response function is not
properly calibrated for the CC-mode. Following the method suggested by
Kaaret et al. (2001) we calibrate the gain offset and scale in {\sl
XSPEC} by comparing the overall CC-mode spectra of \psr\ to that
determined by the ACIS-S3 imaging data. The latter is characterized by
the same model with the absorption column density
$N_H=1.6\times10^{22}$ cm$^{-2}$ and a photon index $\alpha=1.35$
(Camilo et al. 2002).  The resulting gain scale and offset are found
to be 0.90 and -0.18, respectively. Fixing this gain correction and
$N_H$ to the above values, we re-fit the pulsed emission spectrum to
obtain a photon index of $1.2\pm0.2$ (see Figure~6b). The new $\chi^2$ value
is 17.7 for 34 degree-of-freedom, significantly better than without
the gain correction. The pulsed flux
measured in the $2-10$~keV energy band is $1.2\times10^{-12}$ ergs
cm$^{-2}$ s$^{-1}$. When compared to the overall $2-10$~keV flux of
$1.7\times10^{-12}$ ergs cm$^{-2}$ s$^{-1}$ (Camilo et al. 2002), this
implies that $\sim 70\%$ of the total emission from the pulsar is
pulsed, consistent with the estimate in Section 3.1.

\begin{figure*}
\plotone{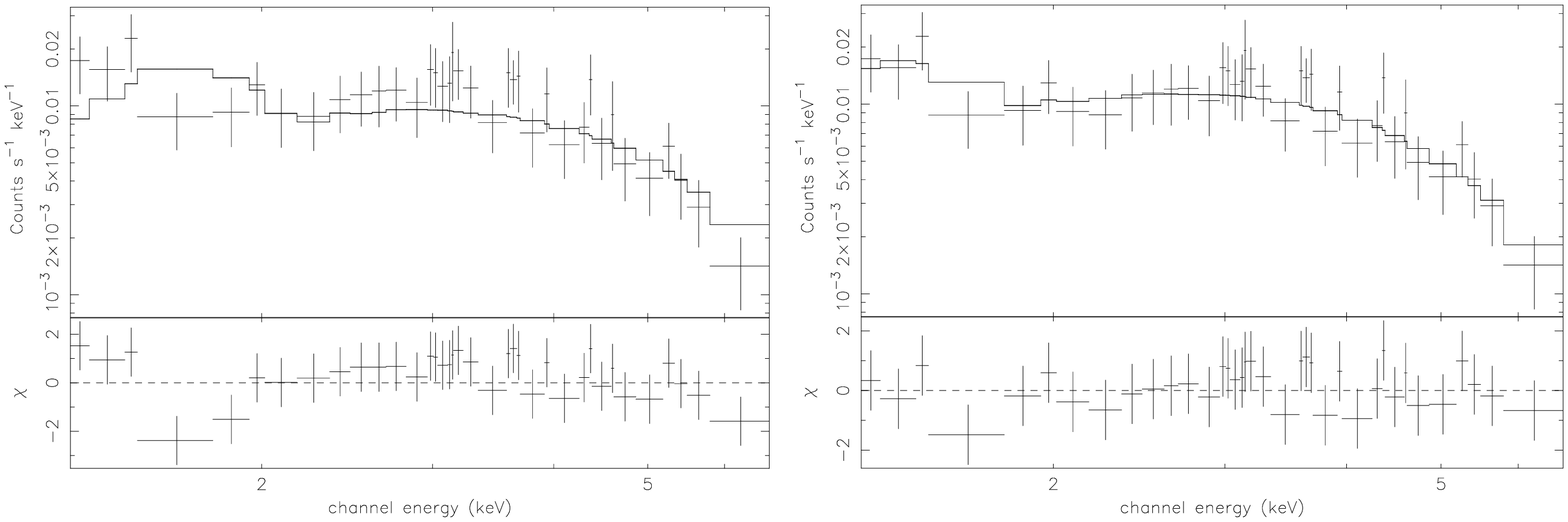}
\caption{The pulsed X-ray spectrum of \psr\ obtained with \chandra\
ACIS-I3 in continuous-clocking mode: {\it Left Panel:} fitting with an
absorbed power-law model with the gain scale and offset fixed as 1 and
0;  {\it Right Panel:} fitting with the same model but with the
gain scale and offset of 0.90 and -0.18.
\label{fig6}}
\end{figure*}

\section{Discussion}
The properties of \psr\ are most similar to those found for other
examples of young, energetic pulsars. The power-law spectral index of
the pulsar emission is consistent with its spin-down energy according
to the empirical law of Gotthelf (2003) for energetic rotation powered
pulsars with $\dot E > 4 \times 10^{36}$~erg~s$^{-1}$. The power law
index is also consistent with that of the pulsed emission, as found
for other high $\dot E$, Crab-like pulsars (Gotthelf 2003). As with
most X-ray detected radio pulsars, the X-ray pulse morphology differs
from that of the radio pulse. The full width at half
maximum (FWHM) of the X-ray pulse is 0.4 phase compared to 0.15 phase
in radio. Notably, the X-ray pulse has a steep rise and slow decline,
whereas the radio pulse is inverted, with a slow rise and steep decay
instead.

The unpulsed component of \psr\ is most likely nonthermal in nature as
the thermal emission from the cooling surface of the neutron star
should be negligible.  According to the standard theoretical cooling
curves, the surface temperature of a 1.4 M$\sun$ neutron star is about
0.13 keV at the age of \psr\ (about 3,000 years; Page 1998). Assuming
a radius of 12 km the neutron star should have an absorbed $0.2-10$
keV flux of $\sim 8 \times 10^{-15}$~erg~cm$^{-2}$~s$^{-1}$, which
accounts for $\sim$ 0.4\% of our detected total $0.2-10$~keV X-ray
flux or 1.4\% of the unpulsed flux. Tennant et al.  (2001) detected
the X-ray emission of the Crab pulsar at its pulse minimum, though
accounting for only a tiny fraction of the total or unpulsed flux.
Tennant et al. (2001) further suggested that this component is
nonthermal.  The unpulsed X-ray emission from \psr\ may be of the same
nature as that of the Crab pulsar.
 
Together with the previous X-ray and radio periods, the three timing
measurements obtained herein provide an opportunity to study the pulsar
period evolution. 
A linear fit to these periods yields a $\dot{P}$ of 7.5116(6)$\times
10^{-13}$~s~s$^{-1}$ with a reduced $\chi^2_{\nu}$ of 3.6 (see Figure~7).
 The large
$\chi^2_{\nu}$ value and the scattered residuals show that the period
of \psr\ evolves in a more complicated than a simply constant spin down.
 The period derivative obtained here is
also significantly ($9\sigma$) different from that obtained by Camilo
et al. (2002). This suggests that \psr\ has experienced periods
of timing noise and/or glitches - not unepxected for a young
pulsar (e.g., Zhang et al. 2001; Wang, et al. 2001; Crawford \& Demia\'nsky
2003). Arzoumanian et al. (1994) 
defined a quantity $\Delta_8$ to 
represent the stability of a pulsar. They found an  empirical relation
between $\Delta_8$ and $\dot{P}$, which predicts a high
$\Delta_8$ of -0.67 for PSR J1930+1852. This value is higher than 
those measured for most ordinary pulsars and is consistent with the
variability in spin-down rate observed for this pulsar.

\begin{figure}
\plotone{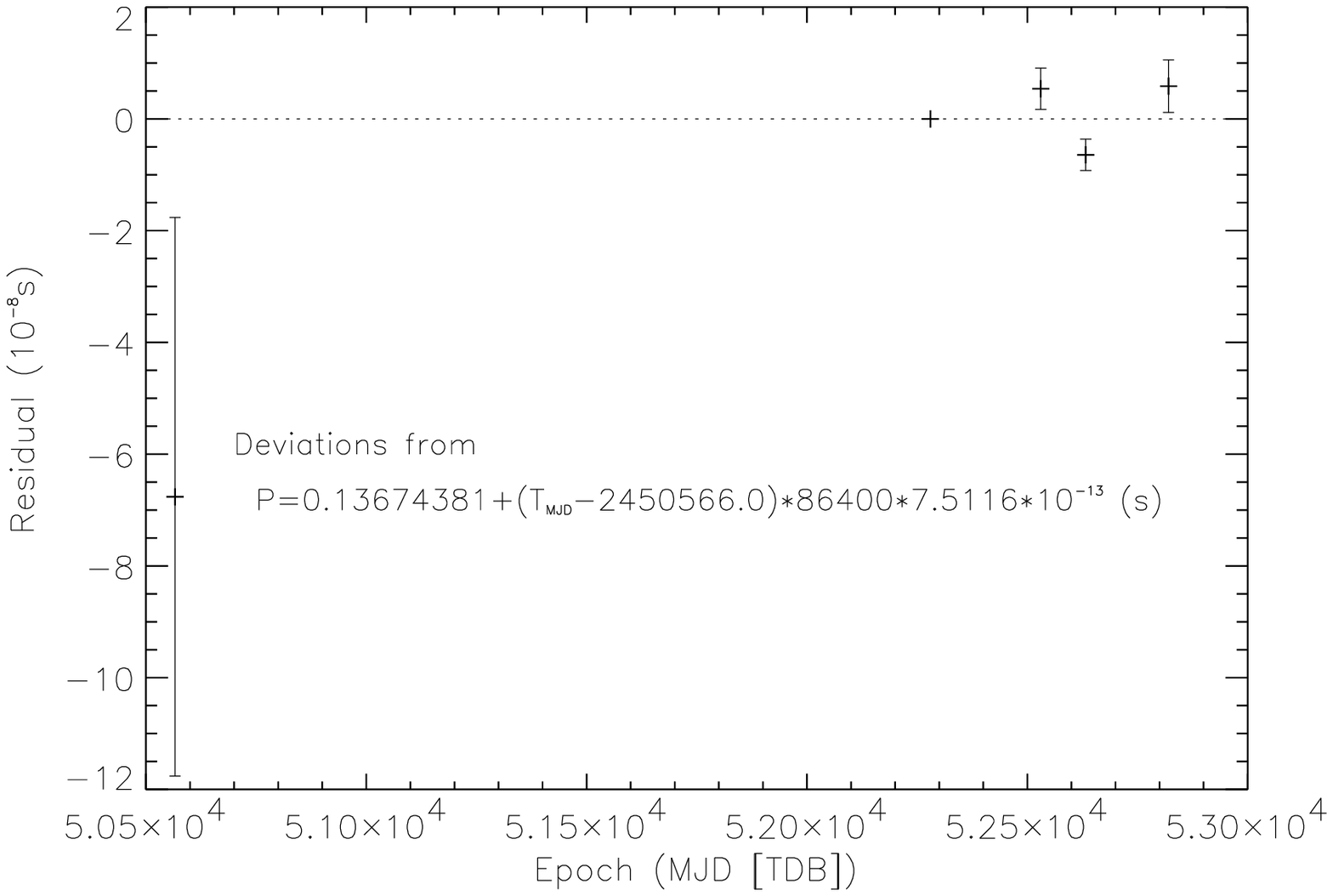}
\caption{The period residuals of PSR J1930+1852 in different epochs.
\label{fig7}}
\end{figure}

Indeed, \psr\ shares other interesting properties with PSR~B0540-69. 
For example, the
pulsed X-ray emission of PSR~B0540-69 has probably a harder spectrum,
with a photon index of $1.83\pm0.13$, than the steady component whose
photon index is ($2.09\pm0.14$; Kaaret et al.  2001), whereas \psr\
also has a harder pulsed emission than the steady emission. Furthermore,
the pulse width of PSR~B0540-69 is about
0.4 and its pulsed fraction $f_p = 71.0\pm5\% $, both nearly identical
to the respective values measured herein for \psr.
Based on these X-ray emission similarities, the X-ray emission
regions of the two pulsars may have the similar overall structures and
viewing geometries.

\acknowledgments The project is partially supported by NASA/SAO/CXC through
grant GO5-6057X. FJL and JLQ also acknowledge support from the National
Natural Science Foundation of China.

\end{document}